\documentclass[aps,twocolumn,superscriptaddress,showpacs]{revtex4-1}
\usepackage[dvips]{graphicx}
\usepackage{latexsym}
\usepackage{amsmath}
\usepackage{amsfonts}
\usepackage{amssymb}
\usepackage{bm}
\usepackage{color}
\begin{document}
\newcommand{\fig}[2]{\includegraphics[width=#1]{#2}}
\newcommand{{\vhf}}{\chi^\text{v}_f}
\newcommand{{\vhd}}{\chi^\text{v}_d}
\newcommand{{\vpd}}{\Delta^\text{v}_d}
\newcommand{{\ved}}{\epsilon^\text{v}_d}
\newcommand{{\vved}}{\varepsilon^\text{v}_d}
\newcommand{\la}{\langle}
\newcommand{\ra}{\rangle}
\newcommand{\dg}{\dagger}
\newcommand{\upa}{\uparrow}
\newcommand{\dna}{\downarrow}
\newcommand{\as}{{\alpha\sigma}}
\newcommand{\hH}{{\hat{\mathcal{H}}}}
\newcommand{\hn}{{\hat{n}}}
\newcommand{\hP}{{\hat{P}}}
\newcommand{{\bk}}{{\bf k}}
\newcommand{{\bq}}{{\bf q}}
\newcommand{{\tr}}{{\rm tr}}
\newcommand{\pprl}{Phys. Rev. Lett. \ }
\newcommand{\pprb}{Phys. Rev. {B}}
\newcommand{\dx}{$d_{x^2}$\ }
\newcommand{\dz}{$d_{z^2}$\ }

\title{Electronic structure and two-band superconductivity in unconventional high-$T_c$ cuprates
Ba$_2$CuO$_{3+\delta}$}

\author{Kun Jiang}
\thanks{These two authors contributed equally}
\affiliation{Beijing National Laboratory for Condensed Matter Physics and Institute of Physics,
	Chinese Academy of Sciences, Beijing 100190, China}
\affiliation{Department of Physics, Boston College, Chestnut Hill, MA 02467, USA}

\author{Congcong Le}
\thanks{These two authors contributed equally}
\affiliation{Kavli Institute of Theoretical Sciences, University of Chinese Academy of Sciences,
Beijing, 100190, China}
\affiliation{Chinese Academy of Sciences Center for Excellence in Topological Quantum Computation, University of Chinese Academy of Sciences, Beijing 100190, China}

\affiliation{Beijing National Laboratory for Condensed Matter Physics and Institute of Physics,
	Chinese Academy of Sciences, Beijing 100190, China}

\author{Yinxiang Li}
\affiliation{Beijing National Laboratory for Condensed Matter Physics and Institute of Physics,
	Chinese Academy of Sciences, Beijing 100190, China}

\author{Shengshan Qin}
\affiliation{Kavli Institute of Theoretical Sciences, University of Chinese Academy of Sciences,
Beijing, 100190, China}
\affiliation{Beijing National Laboratory for Condensed Matter Physics and Institute of Physics,
	Chinese Academy of Sciences, Beijing 100190, China}

\author{Ziqiang Wang}
\email{wangzi@bc.edu}
\affiliation{Department of Physics, Boston College, Chestnut Hill, MA 02467, USA}

\author{Fuchun Zhang}
\email{fuchun@ucas.ac.cn}
\affiliation{Kavli Institute of Theoretical Sciences, University of Chinese Academy of Sciences,
Beijing, 100190, China}
\affiliation{Chinese Academy of Sciences Center for Excellence in Topological Quantum Computation, University of Chinese Academy of Sciences, Beijing 100190, China}

\author{Jiangping Hu}
\email{jphu@iphy.ac.cn}
\affiliation{Beijing National Laboratory for Condensed Matter Physics and Institute of Physics,
	Chinese Academy of Sciences, Beijing 100190, China}
\affiliation{Collaborative Innovation Center of Quantum Matter,
Beijing, China}
\affiliation{Kavli Institute of Theoretical Sciences, University of Chinese Academy of Sciences,
Beijing, 100190, China}

\date{\today}

\begin{abstract}
The recently discovered cuprate superconductor Ba$_2$CuO$_{3+\delta}$ exhibits a high $T_c\simeq73$K at $\delta\simeq0.2$.
The polycrystal grown under high pressure has a structure similar to La$_2$CuO$_4$,
but with dramatically different lattice parameters due to the CuO$_6$ octahedron compression. The crystal field in the compressed Ba$_2$CuO$_4$ leads to an {\em inverted} Cu $3d$ $e_g$ complex with the $d_{x^2-y^2}$ orbital sitting below the $d_{3z^2-r^2}$ and an electronic structure highly unusual compared to the conventional cuprates. We construct a two-orbital Hubbard model for the Cu $d^9$ state at hole doping $x=2\delta$ and study the orbital-dependent strong correlation and superconductivity. For the undoped case at $x=0$, we found that strong correlation drives an orbital-polarized Mott insulating state with the spin-$1/2$ moment of the localized $d_{3z^2-r^2}$ orbital. In contrast to the single-band cuprates where superconductivity is suppressed in the overdoped regime, hole doping the two-orbital Mott insulator
lead to orbital-dependent correlations and the robust spin and orbital exchange interactions produce a high-$T_c$ antiphase $d$-wave superconductor even in the heavily doped regime at $x=0.4$.
We conjecture that Ba$_2$CuO$_{3+\delta}$ realizes mixtures of  such heavily hole-doped superconducting Ba$_2$CuO$_4$ and disordered Ba$_2$CuO$_{3}$ chains in a single-layer or predominately separated bilayer structure.
Our findings suggest that unconventional cuprates with liberated orbitals as doped two-band Mott insulators can be a direction for realizing high-T$_c$ superconductivity with enhanced transition temperature $T_c$.

\typeout{polish abstract}
\end{abstract}


\maketitle

\section{Introduction}
The current understanding of high-$T_c$ cuprate superconductors \cite{bednorz} crucially relies on the crystal field due to the Jahn-Teller distortion of the elongated CuO$_6$ octahedra. The topmost Cu $3d$-electron $e_g$ states split accordingly into well separated lower $d_{3z^2-r^2}$ ($d_{z^2}$) and upper $d_{x^2-y^2}$ ($d_{x^2}$) orbitals. In the parent compound, such as the prototypical single-layer La$_2$CuO$_4$ (La214), the Cu$^{2+}$ is in the $3d^9$ configuration with a fully occupied \dz orbital and an active \dx orbital partially occupied by one electron. The strong correlation produces a spin-${1\over2}$ antiferromagnetic (AF) Mott insulator. Hole doping leads to an effective one-band model of Zhang-Rice singlets formed by the hole in the $d_{x^2}$ orbital and a doped hole in the planar oxygen $2p^5$ orbitals \cite{zhang_rice}. The AF exchange interactions give rise to nodal $d$-wave high-$T_c$ superconductivity in the CuO$_2$ planes \cite{lee}. Such a picture describes the vast majority of the conventional cuprates, where the dormant \dz orbital plays only a minor role \cite{aoki10,aoki12,millis11,devereaux10}.

The recent discovery of the high-$T_c$ superconductor Ba$_2$CuO$_{3+\delta}$ at $\delta\simeq0.2$ \cite{jin} highlights a class of ``unconventional'' cuprates where the different crystal field distributions lead to electronic structures with liberated \dz orbital \cite{jiang,xue}. The polycrystal samples have been synthesized under high pressure in a strongly oxidizing environment.
Extraordinary properties were observed by a combination of magnetization, specific heat, neutron scattering, x-ray diffraction (XRD),  x-ray absorption spectroscopy (XAS), $\mu$ spin-rotation ($\mu$SR) experiments \cite{jin}, and resonant inelastic x-ray scattering (RIXS) \cite{rixs}:
(i) The Ba$_2$CuO$_{3+\delta}$ has an atomic structure similar to La214, but with dramatically different lattice parameters due to octahedral compression, leading to {\em inverted} $d_{z^2}$ and $d_{x^2}$ orbitals; (ii) The extra $O_\delta$ occupy the planar oxygen sites;
(iii) Despite the large hole doping reflected in the Cu $L_3$ XAS and RIXS spectra,
the O K-edge XAS shows spectral weight transfer similar to
La214, indicative of strong Mott-Hubbard correlations \cite{xas,mills}; and remarkably,
(iv) The superconducting (SC) transition temperature $T_c\simeq73$K nearly doubles
that of the La214 family with a SC volume fraction above 30\% at low temperatures.
\begin{figure*}
	\begin{center}
		\fig{7.0in}{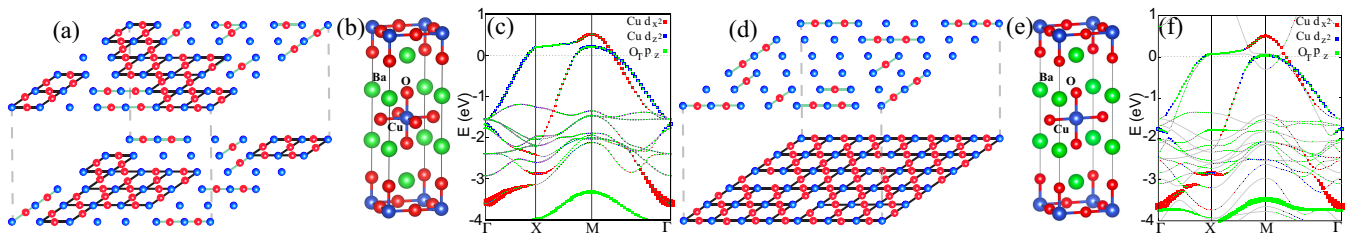}\caption{Mixed structures of (a) Ba214 regions with CuO$_2$ plane and Ba213 regions with short Cu-O chains and substantial oxygen vacancies within a single-layer and (d) bilayer of a mostly Ba214 with CuO$_2$ plane and a disordered Ba213 with short Cu-O chains and substantial oxygen vacancies. (b) and (e): Atomic structures of stoichiometric Ba214 and Ba214/Ba213 bilayer corresponding to (a) and (d). (c) and (f): DFT band structures of compressed Ba214 corresponding to (b) and (e) using structural parameters measured by neutron scattering. Zero marks the Fermi level corresponding to Cu $3d^9$ configuration.
			\label{fig1}}
	\end{center}
	\vskip-0.5cm
\end{figure*}

In this work, we present a theoretical description of the possible electronic structure and the SC state of Ba$_2$CuO$_{3+\delta}$ based on the remarkable experimental findings described above. The focus is to address the outstanding puzzle of high-$T_c$ superconductivity at extremely high hole doping concentration $x=2\delta\sim0.4$ that defies all conventional single-band cuprates where superconductivity is known to be suppressed in the overdoped regime \cite{lee}.
In doing so, we find that the cuprate phenomenology is remarkably enriched due to the liberated \dz orbital in addition to the \dx orbital, and that the doped two-orbital Mott insulator with strongly orbital dependent correlations may provide a mechanism for the emergence of superconductivity with dramatic $T_c$ enhancement. Specifically, we construct a two-orbital Hubbard model describing the low-energy bands of the compressed Ba$_2$CuO$_{4}$
with partially occupied and inverted \dz and \dx bands and study the correlated electronic states.
For the undoped case at $x=0$, we find that strong correlation induces interorbial/interband carrier transfer and  drives an orbital-polarized Mott insulating state with the spin-$1/2$ moment originating from the localized $d_z$ orbital. Hole doping the two-orbital Mott insulator leads to an itinerant state with orbital-dependent strong correlation effects. We find that a novel two-band superconductor with an antiphase $d$-wave gap function due to the robust spin and orbital exchange interactions, suggesting multiband and orbital selectivity are crucial for high-$T_c$ superconductivity in the heavily hole-doped regime of the unconventional cuprate Ba$_2$CuO$_{3+\delta}$.

The rest of the paper is organized as follows. In Section II, the crystal structure and the density functional theory (DFT) calculations are discussed, taking into account the observations by neutron, x-ray scattering, and RIXS experiments. In Section III, we construct a two-orbital tight-binding model for the two low-energy bands in the DFT, and study the correlated electronic states in the Cu $e_g$ complex using the two-orbital Hubbard model at hole doping concentration $x=2\delta$ relative to the 3$d^9$ state. The correlation driven interband carrier transfer and orbital-dependent band renormalization are studied using the strong-coupling Gutzwiller approximation.
We obtain the orbital polarized Mott insulating state at $x=0$ and the correlated paramagnetic state with strongly orbital-dependent correlation effects at $x=0.4$. In Section IV, the strong-coupling Kugel-Khomskii spin-orbital superexchange interactions are derived for the doped two-orbital Mott insulator. We show that the ground state is the novel two-band $d_\pm$-wave superconductor at $x=0.4$ in the renormalized mean field theory. A detailed renormalization group analysis is also presented for the competing instabilities of superconductivity with different pairing symmetries and the incipient density wave orders. Summary and outlook are given in Section V.

\section{
Crystal and electronic structure}

Based on the findings of neutron diffraction and x-ray scattering experiments, the compressed Ba$_2$CuO$_{3+\delta}$ near $\delta\simeq0.2$ is likely to crystalize into mixed structures of Ba$_2$CuO$_{4}$ (Ba214) regions with CuO$_2$ planes, which will be shown to exhibit two-band high-T$_c$ superconductivity, and normal regions of Ba$_2$CuO$_{3}$ (Ba213)
with Cu-O chain planes.
%
One example, shown in Fig.~1(a), has the SC Ba214 containing limited oxygen vacancies embedded in the non-SC disordered Ba213 hosting short Cu-O chains and substantial oxygen vacancies within a single-layer. Intriguingly, evidence from recent RIXS experiments points to a mixed bilayer structure with the disordered Ba213 containing short Cu-O chains and vacancies separated from the SC layer in a unit cell \cite{rixs}. Fig.~1(d) illustrates such a bilayer structure with a close to ideal CuO$_2$ plane of SC Ba214 and a non-SC disordered Cu-O chain plane of Ba213 acting as a charge reservoir layer, similar to YBa$_2$Cu$_3$O$_{7-x}$ \cite{ybco}.
Since Cu$^{4+}$ (3$d^7$) in Ba214 is not a stable ionization state, self-doping takes place with respect to the Cu$^{2+}$ in Ba213 both within the plane as in Fig.~1(a) and across the bilayer as in Fig.~1(d).
The absence of apparent charge disproportionation in $\mu$SR \cite{jin} suggests equal valence of the Cu atoms in the Ba213 and Ba214 regions. This leads to a heavily overdoped average hole concentration $x=2\delta\simeq40\%$ relative to the Cu 3$d^9$ configuration, independent of specific realizations of the mixed structure.

Using the structural parameters measured by neutron scattering summarized in Table 1 in Appendix A, we carried out DFT calculations for the stoichiometric structures of  the compressed Ba214 (Fig. 1(b)) and Ba213/Ba214 bilayer (Fig.~1(e)), as well as Ba213 using the Vienna {\em ab initio} simulation package (VASP) \cite{Kresse1993,Kresse1996,Kresse1996b}. The band structures are calculated in the generalized gradient approximation \cite{Perdew1996,Monkhorst1976}.
The details of the DFT calculations are given in Appendix A.
The orbital-resolved electronic structure of Ba214 is shown in Fig.~1(c). There are two bands crossing the Fermi level, involving predominately the Cu $3$\dx and $3$\dz orbitals. Since the compressed CuO$_6$ octahedron has a shorter distance connecting the Cu to the apical oxygens than to the planar oxygens (Table 1, Appendix A), the \dz orbital couples strongly to the apical oxygen $p_z$ orbital. The {\em inverted} crystal field pushes the energy of the \dz orbital to lie $\sim0.93$eV above the \dx orbital. Thus the atomic $3d^9$ configuration has one-electron in the \dz orbital, while the \dx orbital is fully occupied by two electrons, in contrast to the conventional cuprates.

Upon crystallization, although the smaller hopping integrals of the out of plane \dz orbital produce a narrow band, the \dx orbital generates a much wider band through the larger hopping integrals via the planar oxygens, such that the two bands overlap near the Fermi level.
For the Ba213/Ba214 bilayer structure, we find that the orbital-resolved band dispersions in the Ba214 plane, shown in Fig.~1(f), are remarkably similar to those in the compressed single-layer Ba214 (Fig.~1(c)). The projected electronic structure in the Ba213 plane 
is, on the other hand, similar to that of the bulk single-layer Ba213
as discussed in Appendix A, with a single 1D band near the Fermi level \cite{taoxiang}.
The primary $d$-electron content is $d_{x^2-z^2}$, a linear combination of the \dx and \dz orbitals, for the ideal Cu-O chains along the $x$-direction. The $40\%$ hole-doping of the single-band and the strong disorder effects associated with the oxygen vacancies and short Cu-O chains shown in Figs.~1(a,d) are expected to lead to disordered metallic Ba213 regions.

We thus study the compressed Ba214 as an unconventional {cuprate} where both \dx and \dz orbitals in the $e_g$ sector contribute to superconductivity. Although multi-orbital superconductivity has been studied extensively for the $t_{2g}$ electrons in the iron pnictides and chalcogenides superconductors \cite{mazin,hu,frg,hirshfeld-scalapino,chubukov08},
this is rare for the $e_g$ electrons in the cuprates. We will show that 
a new two-band $d$-wave superconductor with antiphase pairing gaps on the two Fermi surfaces (FSs) emerges from the multiorbital correlated electronic structure at the high doping $x=0.4$ achieved experimentally.

\section{Two-orbital Hubbard model and orbital-dependent correlated states}

To this end, we construct an effective two-orbital model on the Cu square lattice $H=H_t + H_I$, where $H_t$ describes the DFT band structure near the Fermi level and $H_I$ the electron correlations. The model constitutes a generalized Zhang-Rice singlet construction where the charge degrees of freedom on the oxygen sites have been integrated out, and the Cu$^{3+}$ in the 3d$^8$ configuration includes the spin singlets of a hole (3d$^9$) on Cu-site and a hole (2p$^5$) on its neighbouring O-sites with compatible symmetries to the \dx and $d_z$ orbitals \cite{zhang_rice}.
\begin{figure}
	\begin{center}
		\fig{3.4in}{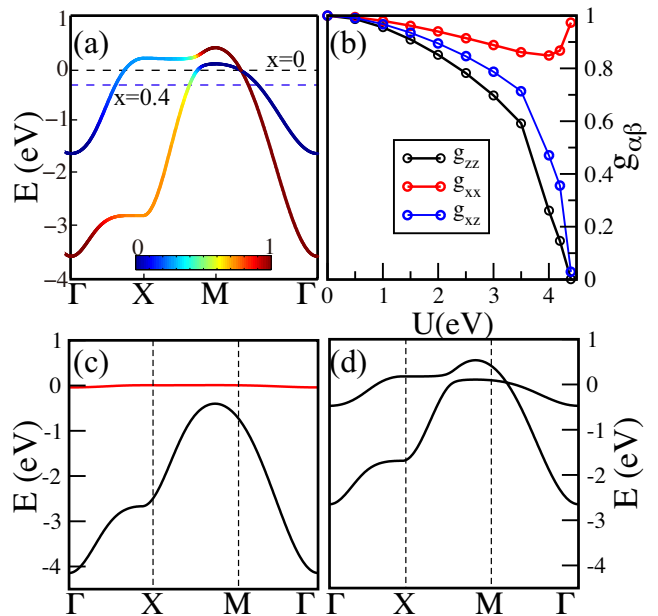}\caption{ (a) Band dispersion in the TB model with color coded orbital content: \dx orbital (red and 1) and  \dz orbital (blue and 0). The Fermi levels are marked by black ($x=0$) and blue ($x=0.4$) dashed lines. (b) The hopping renormalization factor $g_{\alpha\beta}$ as a function of $U$ at $x=0$, showing a Brinkman-Rice transition at $U_c\simeq4.4eV$. (c) Correlated band dispersion at $x=0$ and $U=4.4eV$.
(d) Correlated band structure at $x=0.4$ and $U=7$eV.}
	\end{center}
	\vskip-0.5cm
\end{figure}

\subsection{Two-orbital tight-binding model}

Denoting a spin-$\sigma$ electron in the effective \dx and \dz like orbitals by $d_{\alpha\sigma}$ with $\alpha=x,z$, the tight-binding (TB) model is given by
\begin{eqnarray}
H_{t}&=&\sum_{k\alpha\beta\sigma}\varepsilon_k^{\alpha\beta}
d_{k\alpha\sigma}^\dagger d_{k\beta\sigma} + \sum_{k\sigma} e_\alpha d_{k\alpha\sigma}^\dagger d_{k\alpha\sigma},
\label{ht}
\end{eqnarray}
where $e_\alpha$ denotes the crystal field energy of each orbital. The lattice structure of Ba214 belongs to the $D_{4h}$ point group. The intra and interorbital hopping can be expressed in terms of the lattice harmonics of different symmetry. Specifically, $\gamma_k=\cos k_x+\cos k_y$, $\alpha_k=\cos k_x\cos k_y$, and $\gamma_k^\prime=\cos2k_x+\cos2k_y$ in the $A_{1g}$ channel and $\beta_k=\cos k_x-\cos k_y$ and $\beta_k^{\prime}=\cos2 k_x-\cos 2k_y$ in the $B_{1g}$ channel, for up to third nearest neighbors. The corresponding expressions for the hopping energies in Eq.~(\ref{ht}) are thus given by
\begin{eqnarray}
\varepsilon_k^{\alpha\alpha}&=&-2t_\alpha \gamma_k-4t_\alpha^\prime \alpha_k-2t_\alpha^{\prime\prime}\gamma^\prime_k \\
\varepsilon_k^{xz}&=&2t_{xz} \beta_k+2t_{xz}^{\prime\prime}\beta^\prime_k,
\label{epsilons}
\end{eqnarray}
where the values of the hopping parameters are given in Table 2 together with $e_\alpha$ in Appendix B.

The band structure of the two-orbital TB model is shown in Fig.~2(a). It faithfully represents the low energy DFT band structures in Figs.~1(c,f), including the orbital content, as a two-band system with a bandwidth $W\simeq4$eV. Since the interorbital mixing $\varepsilon_k^{xz}$ has $B_{1g}$ symmetry, the band structure along the
$\Gamma\to M$ direction is orbital diagonal. Moreover, the interband mixing is also weak along the $\Gamma\to X$ direction because of the large band energy separation. However, the two bands hybridize strongly when the band energies are close, as seen along the $X\to M$ direction.
The undoped $d^9$ configuration at $x=0$ has $n_e=3$ electrons with the bare orbital occupations $n_z^0\simeq1.3$ and $n_x^0\simeq1.7$. For
the experimental doping level corresponding to $x=0.4$,  $n_e=2.6$ and the bare occupations change to $n_z^0=1.04$ and $n_x^0=1.56$. There are two Fermi surfaces (FSs), one electron-like of dominant \dz character around the $\Gamma$ point at the center and a much smaller hole-like \dx FS around the $M$ point in the Brillouin zone.
We next turn to the correlation effects beyond the band description.

\subsection{Hubbard correlations and Gutzwiller approximation}

The correlation part of the Hamiltonian $H_I$ follows the two-orbital Hubbard model for the $e_g$ complex \cite{castellani,georges,kk},
\begin{align}
H_I&=U\sum_{i,\alpha}\hn_{i\alpha\upa}\hn_{i\alpha\dna}
+\left(U'-{1\over 2}J_H\right)\sum_{i,\alpha<\beta}\hn_{i\alpha}\hn_{i\beta}
\label{hi} \\
&-J_H\sum_{i,\alpha\neq\beta}{\bf S}_{i\alpha}\cdot {\bf S}_{i\beta}
+J_H\sum_{i,\alpha\neq\beta}d^\dg_{i\alpha\upa}
d^\dg_{i\alpha\dna}d_{i\beta\dna}d_{i\beta\upa},
\nonumber
\end{align}
where the intra and interorbital repulsion $U$ and $U^\prime$ are related by the Hund's rule coupling $J_H$
through $U=U'+2J_H$. Since $J_H$ is small for the $e_g$ electrons, it is set as $J_H=0.1U$  and does not affect our results.

We next study the strong correlation effects using the multiorbital Gutzwiller projection method \cite{gebhard98,lechermann,sen}: $H=H_t+H_I\to H_G=P_GH_tP_G$, where $P_G$ is the finite-$U$ Gutzwiller projection operator that reduces the statistical weight of the Fock states with multiple occupations.
The projection can be conveniently implemented using the Gutzwiller approximation
\cite{gebhard98,lechermann} developed to study the multiorbital cobaltates \cite{cobaltate}, Fe-pnictides \cite{sen}, and the monolayer CuO$_2$ grown on Bi$_2$Sr$_2$CaCu$_2$O$_{8+\delta}$ substrate \cite{jiang}. In this approach, the Gutzwiller projected Hamiltonian is expressed as
\begin{equation}
H_{G}=\sum_{k\alpha\beta\sigma}g_{\alpha\beta}^\sigma\varepsilon_k^{\alpha\beta}
d_{k\alpha\sigma}^\dagger d_{k\beta\sigma} + \sum_{k\sigma}(e_\alpha +\lambda_\alpha^\sigma)d_{k\alpha\sigma}^\dagger d_{k\alpha\sigma}.
\label{ght}
\end{equation}
The strong correlation effects are described by the orbital dependent hopping renormalization $g_{\alpha\beta}^\sigma$ and the renormalized crystal field $\lambda_\alpha^\sigma$, which must be determined self-consistently for a given electron density.

\subsection{Orbital-polarized Mott insulator as $x=0$}

We first study the correlation effects in the undoped case at $x=0$. The band renormalization factors $g_{\alpha\beta}$  are calculated self-consistently and plotted in Fig.~2(b) as a function of the Hubbard $U$. Clearly, these band narrowing factors are strongly orbital dependent.
As $U$ is increased, the interorbital/interband carrier transfer ensues \cite{cobaltate} and drives the \dz band toward half-filling with $n_z=1$. As a result, the band narrowing factors $g_{zz}$ and $g_{xz}$ reduce with increasing $U$ and vanish at $U_c\simeq4.4eV$, indicating an orbital-selective Mott transition of the Brinkman-Rice type \cite{brinkman-rice}. The renormalized band dispersions in the paramagnetic phase are plotted in Fig.~2(c) near the transition,
showing a half-filled flat \dz band of localized spin-1/2 moment and a wide, completely filled \dx band below the Fermi level. Thus, the ground state is an orbital-polarized Mott insulator for $U>U_c$ with the insulating gap and
antiferromagnetic ordered moments originating from the electrons in the \dz orbital.


\subsection{Orbital-dependent correlations - normal state at $x=0.4$}

We next study the correlation effects in the hole doped two-orbital Mott insulator for $U>U_c$. First, we focus on the normal state at large hole doping $x=0.4$ relevant for the experiments. For the conventional cuprates, the correlation effects are significantly weakened in the overdoped regime, consistent with doping a single-band Mott insulator \cite{lee}. The situation changes significantly in the case of doping two-orbital Mott insulator. To stay on the Mott insulating side, we choose a large on-site replusion $U=7$eV, which corresponds to a correlation to bandwidth ratio of $U/W\simeq1.8$ typically considered for the cuprates \cite{dagotto,lee}. Fig. ~2(d) displays the Gutzwiller renormalized band dispersions obtained self-consistently for an average electron number $n_e=2.6$ in the two $e_g$ orbitals, with the corresponding FSs shown in Fig.~3(a).
Interestingly, the calculated orbital occupations $n_z=1.02$ and $n_x=1.58$ are close to their bare values discussed in section III.B, reflecting the tendency of the correlations to keep the \dz orbital close to half-filling. As a result, the band renormalization effect remains strong and orbital-dependent. The strongly correlated \dz band narrows significantly by a factor $g_{zz}=0.33$, while the \dx band only narrows by a factor $g_{xx}=0.87$. Consequently, the orbital-dependent correlation effect in the multiorbital doped Mott insulator is responsible for the heavily hole doped compressed Ba214 to remain as a strongly correlated Mott-Hubbard system.

\section{Two-band superconductivity}

It is thus conceivable that a two-band superconductor can emerge from the spin and orbital superexchange interactions of the doped two-orbital Mott insulator, even at the very high doping level of $x=0.4$. The spin-orbital superexchange interactions are of the Kugel-Khomskii type \cite{kk,castellani}, which have been recently derived for the two-orbital Mott insulator for the 3$d$ Cu $e_g$ electrons \cite{jiang}. In the spin-orbital basis $(d_{x\uparrow}, d_{x\downarrow},d_{z\uparrow}, d_{z\downarrow})^{T}$, the fermion bilinears at each site can be represented as a tensor product $T_i^\mu S_i^\nu$, where $S_i^\mu$ and $T_i^\mu$, $\mu=0,x,y,z$ are the identity and Pauli matrices divided by 2, acting in the spin and orbital sectors, respectively.
For example, $d_{ix\downarrow}^\dagger d_{iz\uparrow}=T_{i}^{+}S_{i}^{-}$, where $T_{i}^{\pm}=T_i^x\pm iT_i^y$ and $S_{i}^{\pm}=S_i^x\pm iS_i^y$. The spin-orbital superexchange interactions can thus be derived and written as
\begin{eqnarray}
H_{\rm J-K}=\sum_{\langle ij\rangle}\biggl[ J{\bf S}_{i}\cdot{\bf S}_j
&+&\sum_{\mu\nu} I_{\mu\nu} T_i^\mu T_j^\nu
\label{hs} \\
&+&\sum_{\mu\nu} K_{\mu\nu}({\bf S}_{i}\cdot{\bf S}_j)(T_i^\mu T_j^\nu)\biggr]
\nonumber
\end{eqnarray}
where the $J$-term is the SU(2) invariant Heisenberg spin exchange coupling, while the terms proportional $I_{\mu\nu}$ and $K_{\mu\nu}$
describe the anisotropic orbital and spin-orbital entangled superexchange interactions, respectively, since the orbital rotation symmetry is broken by the lattice in the hopping Hamiltonian $H_t$.

\subsection{Renormalized meanfield theory and antiphase $d$-wave superconducting state}

In Ref.~\cite{jiang}, it was shown that despite the orbital order in $T_i^z$, i.e. $\langle T_i^z\rangle\ne0$, induced by the crystal field, the transverse orbital fluctuations associated with $T_i^\pm$ contribute to SC pairing. The possible SC state can be studied using the renormalized meanfield theory commonly applied to studying the pure spin superexchange interaction induced superconductivity in conventional cuprates described by the single-band $t$-$J$ model \cite{lee}.
Including all spin-singlet pairing order parameters
\begin{equation}
\Delta_{ij}^{\alpha\beta\dagger}=d_{i\alpha\uparrow}^\dagger d_{j\beta\downarrow}^\dagger-d_{i\alpha\downarrow}^\dagger  d_{j\beta\uparrow}^\dagger,
 \label{pairingfield}
\end{equation}
in the spin and spin-orbit entangled quadruple exchange interactions in Eq.~(\ref{hs}), we arrive at the effective Hamiltonian describing the SC ground states in the strongly correlated two-orbital model
\begin{eqnarray}
H=P_GH_tP_G&-&\sum_{\langle ij\rangle}\biggl[{J_s\over2}\sum_{\alpha\beta}\Delta_{ij}^{\alpha\beta\dagger} \Delta_{ij}^{\alpha\beta}
\nonumber  \\
&+&{K\over2}\sum_{\alpha\neq\beta}(\Delta_{ij}^{\alpha\alpha\dagger}\Delta_{ij}^{\beta\beta}+\Delta_{ij}^{\alpha\beta\dagger}
\Delta_{ij}^{\beta\alpha})\biggr]
\label{htjk} \\
\nonumber
\end{eqnarray}
The couplings $(J_s,K)$, which are explicit but complicated functions of $t_{\alpha\beta}$, $U$, and $J_H$ \cite{jiang}, will be considered as phenomenological parameters in our effective theory. Evidence from RIXS experiments shows that the spin exchange interaction has an average value $\sim150$meV in Ba$_2$CuO$_{3.2}$, which can be as high as $180$meV \cite{rixs}, similar to the conventional cuprates \cite{lee}. We thus set $J_s=200$meV and treat the orbital exchange coupling $K$ as a parameter. The emergent SC state and its pairing symmetry can be obtained by calculating
the expectation values of the nearest neighbor pairing fields
\begin{eqnarray}
\langle \Delta_{ij}^{\alpha\beta}\rangle&=&\frac{1}{N_s}
\sum_{\mathbf{k},\alpha\beta}\Delta_{\alpha\beta}
b_{\alpha\beta}({k})e^{i\mathbf{k}\cdot({\bf r}_i-{\bf r}_j)},
\label{pairing}
\end{eqnarray}
self-consistently from Eq.~(\ref{htjk}) in the Gutzwiller approximation. Here $N_s$ is the number of lattice sites and $b_{\alpha\beta}({k})$ the symmetry form factors in the $D_{4h}$ point group of the crystal. Specifically, $b_{\alpha\alpha}(k)=\gamma_k$ and $b_{xz}(k)=\beta_k$ in the $A_{1g}$ channel, and $b_{\alpha\alpha}(k)=\beta_k$ and $b_{xz}(k)=\gamma_k$ in the $B_{1g}$ channel.
\begin{figure}
	\begin{center}
		\fig{3.4in}{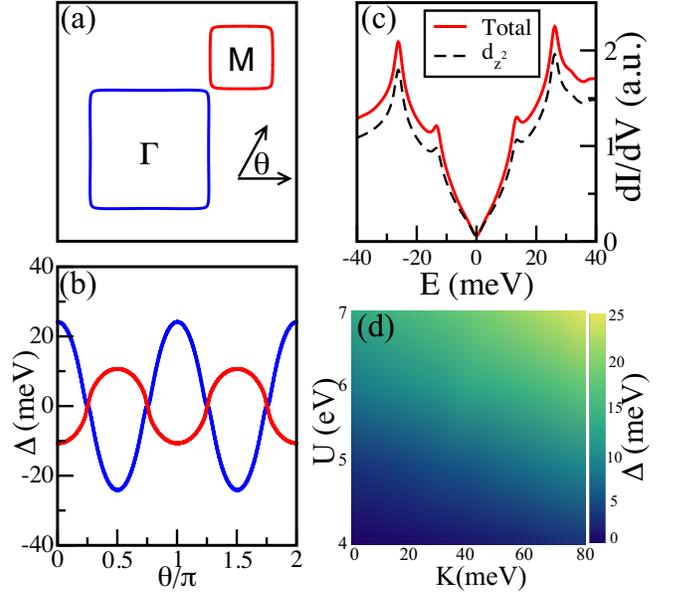}\caption{(a) Normal state FSs at $x=0.4$. (b) Anisotropic pairing energy gaps as a function of angle $\theta$ depicted in (a) along the two FSs around $\Gamma$ (blue) and $M$ (red), showing two antiphase $d$-wave gap functions.
(c) Total (red line) and \dz orbital contribution (black dashed line) to local tunneling density of states, showing two $d$-wave gaps with coherent peaks.
A thermal broadening of 0.5meV is used. (d) Variations of the larger $d$-wave gap as a function of $U$ and $K$.}
	\end{center}
	\vskip-0.5cm
	\label{scorder}
\end{figure}

Remarkably, the ground state is found to be a prominent superconductor with $B_{1g}$ symmetry at the high doping concentration $x=0.4$ relevant for the experiment. In Fig.~3(a) and 3(b), the obtained gap functions at $K=80$meV are plotted along the two FSs. They exhibit a two-band antiphase $d$-wave structure with $4$ gap nodes on each FS and an overall sign-change between the two FSs. This is a strong coupling $d_\pm$-wave analogy of the proposed $s_\pm$ gap function in Fe-based superconductors. This new mechanism of doping an orbital-selective Mott insulator in the two-band unconventional cuprate is crucial for the emergent high-$T_c$ SC state in the highly overdoped region. The close to half-filled, strongly correlated \dz band is responsible for the large pairing gap ($\sim25$meV). The highly overdoped wider \dx band develops a smaller ($\sim9$meV) antiphase gap and boosts the superfluid density for the phase coherence. To probe the basic spectroscopic properties of the novel two-band $d$-wave superconductor, we calculate the orbital resolved local tunneling density of states,
\begin{equation}
N_\alpha(\omega)=\sum_{k\sigma}{\rm Im}\int_0^\beta e^{{i\omega}\tau}\langle{\rm T}_\tau d_{k\alpha\sigma}(\tau)d_{k\alpha\sigma}^\dagger(0)\rangle.
\end{equation}
The total local density of states $N(\omega)=N_x(\omega)+N_z(\omega)$ and the $N_z(\omega)$ are plotted in Fig.~3(c), showing the mixing of two $d$-wave gap structures with the dominant spectral weight coming from the \dz band.

Our results show that the ground state at $x=0.4$ is always the antiphase $d$-wave SC state in the explored parameter space of the Hubbard $U$ and the orbital exchange interaction $K$. Fig.~3(d) shows the map of the larger $d$-wave gap magnitude in the $(K,U)$ plane. The lower limit of $U$ is chosen to be close to $U_c$ of the Mott transition, so that our model describes the doped two-orbital Mott insulator. Note that the gap is appreciable for $U >5.5$eV even if $K=0$. Such a strongly correlated two-band superconductor can provide both a sizable pairing amplitude and a substantial superfluid density, which together control the high-$T_c$ superconductivity in doped Mott insulators \cite{lee}. It is thus conceivable that this mechanism provides an explanation for the nearly doubled $T_c$ compared the isostructural single-band La214.

\subsection{Renormalization group analysis for superconductivity and density wave orders}

The strong-coupling results can be further supported by direct weak-coupling renormalization group (RG) studies of the {two-orbital} Hubbard model defined in Eqs. (1) and (2). This will also allow us to study the competition among superconducting instabilities of different pairing symmetries as well as instabilities toward spin or charge density waves in a unbiased manner. Indeed, the model at $x=0.4$ has two rounded squarish FSs centered at $\Gamma$ and $M$ points of the Brillouin zone as shown in Fig.~3(a). The nearly parallel sections of the FSs raise the issue of competing instabilities involving incommensurate density waves in addition to superconductivity. In metals with nested FS sections, incommensurate spin density wave (SDW) or charge density wave (CDW) can become the ground state. However, in the presence of pairing interactions, the ground state is usually a superconductor since it is more effective at gapping out the entire FS by developing the SC gap.
For example, although an SDW instability is present for a single idealized square FS with perfect nesting, careful renormalization group (RG) studies have shown that the rounding of its corners causes $d$-wave superconductivity to be the leading instability. The ground state is therefore a superconductor and the SDW fluctuation only plays a role at intermediate energy scales or at elevated temperatures \cite{dzyaloshinskii,doucot}. 

For our highly overdoped two-orbital Mott insulator, the two rounded square FSs in Fig.~3(a) are very different in size, leading to two very different nesting vectors $Q_1$ and $Q_2$ connecting the flat sections of the FSs. As a result, the incommensurate SDW and CDW with either $Q_1$ or $Q_2$ are further frustrated since they cannot gap out all the FSs and are thus energetically unfavorable than the SC states that can gap out both FSs entirely. The SC state usually emerges with the $d$-wave symmetry because the large AF fluctuations near $(\pi,\pi)$ mediate a repulsive interaction in the pairing channel. Hence the nontrivial solution of the gap function to the BCS gap equation would change sign along the FS, gapping out the latter except for a set of measure-zero $d$-wave nodes.
In this subsection, we elaborate on this point by performing an RG analysis \cite{furukawa,chubukov08,dzyaloshinskii,furukawa,gology,yao,doucot} for the leading instability of the two FSs. The RG results concretely demonstrate that, even for two perfectly square FSs of very different areas, the leading instability is indeed toward the antiphase $d$-wave superconductivity and the density waves are subleading, which further support the prediction of the strong coupling theory.
Note that the situation changes when the areas of the two FSs are comparable and close to being nested by the wave vector $(\pi,\pi)$.  Nodeless $s$-wave pairing can become the dominate SC state, where the sign of the $s_\pm$ gap function changes across the two FSs. This situation happens at much higher doping where the Cu is close to the $d^8$ configuration and will be discussed in the next subsection.

To this end, we approximate the two FSs in Fig.~3(a) by two perfectly square FSs around $\Gamma$ and $M$ points, respectively, as shown in Fig.~4(a). The 8 flat sections of the FSs are represented by 8 patches in the RG \cite{furukawa,chubukov08} marked by the solid and dashed circles on the FS around $\Gamma$ and by the solid and dashed triangles on the FS around $M$. On each FS, the two parallel sections are related by inversion, or inversion followed by translation by a reciprocal lattice vector. They introduce the nesting vectors $(Q_1, 0)$, $(0, Q_1)$, and $(Q_2, 0)$, $(0, Q_2)$, respectively, with $Q_1 > Q_2$.
Hence, besides the logarithmical divergent Cooper susceptibility $\chi^{pp}_0(\mathbf{q})$ at $\mathbf{q}=(0,0)$, the particle-hole susceptibilities $\chi^{ph}_0(\mathbf{q})$ at the nesting vectors $\mathbf{q}=(Q_{1,2},0)$ and $\mathbf{q}=(0,Q_{1,2})$ also diverge.
\begin{figure}
	\begin{center}
		\fig{3.4in}{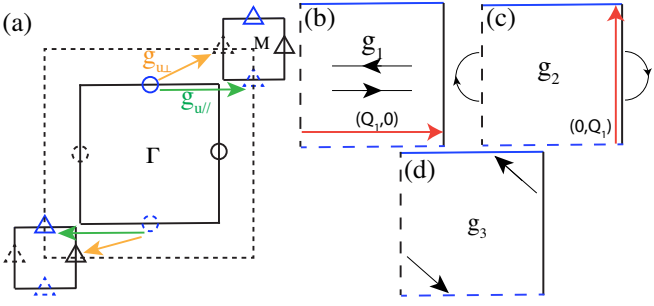}\caption{(a) The two FSs (solid squares) inside the first BZ (dashed square). Circles and triangles denote the 8 patches, each representing one flat section of the FS. The inter-FS coupling $g_{u\parallel}$  (green arrows) describes the pair scattering between two same colored patches (circles and triangles), while $g_{u\perp}$ (orange arrows) describes the pair scattering between two different colored patches.
			(b)The intra-FS coupling $g_1$ describes the backward scattering of electrons between solid and dashed patches. The red line indicates the $(Q_1,0)$ nesting vector. (b) $g_2$ describes the forward scattering of electrons on the solid and dashed patches. The red line indicates the the (0,$Q_1$) nesting vector. (c) $g_3$ describes the scattering of electrons between black and blue patches. }
	\end{center}
	\label{scorder}
\end{figure}

The couplings in the 8 patch model can be classified into the intra-FS and inter-FS couplings. For the intra-FS scattering, there are three relevant coupling constants $g_{1}$, $g_2$, and $g_3$ as shown in Fig.~4(b)-(d). The remaining couplings turn out to be irrelevant here \cite{yao}. For the inter-FS scattering, since $Q_1\neq Q_2$, the coupling constants in the particle-hole sector related to SDW and CDW associated with each FS do not couple. Thus, only the $\mathbf{q}=0$ Copper channel is divergent. The relevant coupling is the pair scattering $g_{u}$ between the two FSs \cite{chubukov08}, which is an Umklapp process with momentum transfer $(2\pi, 2\pi)$ of the reciprocal lattice vector. We decompose $g_{u}$ into $g_{u\parallel}$ (green arrows in Fig.~4(a)) between two parallel patches and $g_{u\perp}$ (orange arrows in Fig.~4(a)) for two patches that are perpendicular. This allows us to account for both the in-phase and antiphase pairing gap functions on the two FSs. The effective coupling constants in the CDW, SDW, antiphase $d$-wave paring (dSC$^\pm$) and $s$-wave paring (sSC$^\pm$), in-phase $d$-wave pairing (dSC$^{++}$) and $s$-wave pairing (sSC$^{++}$) channels can all be expressed in terms of the couplings $g_i$. We obtain
\begin{eqnarray}
	\Gamma_{CDW}^{(Q_1,0)}&=&g_2-2g_1
	\nonumber\\
	\Gamma_{SDW}^{(Q_1,0)}&=&g_2
	\nonumber\\
	\Gamma_{CDW}^{(Q_2,0)}&=&\Gamma_{CDW}^{(0,Q_2)}=
\Gamma_{CDW}^{(0,Q_1)}=\Gamma_{CDW}^{(Q_1,0)}
	\nonumber\\
	\Gamma_{SDW}^{(Q_2,0)}&=&\Gamma_{SDW}^{(0,Q_2)}
=\Gamma_{SDW}^{(0,Q_1)}=\Gamma_{SDW}^{(Q_1,0)}
\label{gamma-densitywave}
\end{eqnarray}
in the particle-hole density wave channels, and
\begin{eqnarray}
	\Gamma_{dSC}^{\pm}&=&g_1+g_2-2g_3-(g_{u\parallel}-g_{u\perp})
\nonumber\\
	\Gamma_{dSC}^{++}&=&g_1+g_2-2g_3+(g_{u\parallel}-g_{u\perp})
\nonumber \\
	\Gamma_{sSC}^{\pm}&=&g_1+g_2+2g_3-(g_{u\parallel}+g_{u\perp})
\nonumber \\
	\Gamma_{sSC}^{++}&=&g_1+g_2+2g_3+(g_{u\parallel}+g_{u\perp})
\label{gamma-pairing}
\end{eqnarray}
in the particle-particle pairing channels.

Evaluating the corresponding Feynman diagrams involving the divergent susceptibilities $\chi^{pp}_0$ and $\chi^{ph}_0$, we obtain the coupled RG flow equations for the coupling constants to leading order \cite{furukawa,chubukov08},
\begin{eqnarray}
	\frac{dg_1}{dy}&=&-2g_1 g_2-2g_3^2+2\alpha g_1 (g_2-g_1)-2(g_{u\parallel}^2+g_{u\perp}^2) \nonumber \\
	\frac{dg_2}{dy}&=&-(g_1^2+g_2^2 )-2g_3^2+\alpha g_2^2-2(g_{u\parallel}^2+g_{u\perp}^2) \nonumber  \\
	\frac{dg_3}{dy}&=&-2g_3 (g_1+g_2)-4g_{u\parallel}g_{u\perp} \nonumber  \\
	\frac{dg_{u\parallel}}{dy}&=&-2(g_1+g_2) g_{u\parallel}-4g_3g_{u\perp}  \nonumber  \\
	\frac{dg_{u\perp}}{dy}&=&-2(g_1+g_2)g_{u\perp}-4g_3g_{u\parallel}
\label{flow}
\end{eqnarray}
where $y=\chi^{pp}_0$ is the RG flow parameter, and $\alpha=\chi^{ph}_0/\chi^{pp}_0$ is the ratio of the bare susceptibilities in the Cooper channel and particle-hole channel, which is generally smaller than unity. The instability of the system can be obtained by integrating the RG equations in Eqs.~(\ref{flow}) for the flow of the coupling constants. The calculated RG flow of the effective coupling constants $\Gamma_\ell$ for different density waves in Eqs.~(\ref{gamma-densitywave}) and for different pairings in Eqs.~(\ref{gamma-pairing})
are shown in Fig.~5 for generic repulsive interactions with initial values $g_{1-3}>0$. A positively divergent effective coupling constant in the density wave channel and a negatively divergent effective coupling constant in the pairing channel indicate the instability of the FSs. The instability temperature $T_\ell$ for developing each order can be defined by the condition of a nonzero solution in the corresponding linearized equation for the order parameter, leading to $T_\ell\sim \varepsilon_Fe^{-{1\over\vert\Gamma_\ell\vert}}$ \cite{chubukov08}. The leading instability, which determines the true instability of the system, is given by the channel that has the highest instability temperature, i.e. the most divergent interaction $\Gamma_\ell$ under the RG flow.
\begin{figure}
	\begin{center}
		\fig{3.4in}{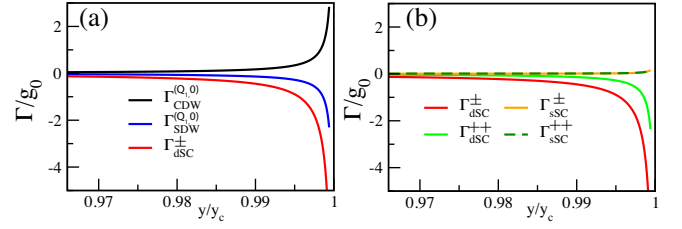}\caption{ The RG flow of the effective couplings ($\Gamma$'s) normalized by a scaling factor $g_0$. (a) Comparison of the RG flow of antiphase $d$-wave superconductivity ($\Gamma_{dSC}^{\pm}$) (red) to those associated with density waves: the SDW ($\Gamma_{SDW}^{(Q_1,0)}$) (blue) and CDW ($\Gamma_{CDW}^{(Q_1,0)}$) (black). (b) Comparison of the RG flow of antiphase $d$-wave superconductivity ($\Gamma_{dSC}^{\pm}$) (red) to other superconductivity channels: inphase $d$-wave superconductivity ($\Gamma_{dSC}^{++}$) (green), antiphase $s$-wave superconductivity ($\Gamma_{sSC}^{\pm}$) (orange), and inphase $s$-wave superconductivity ($\Gamma_{sSC}^{++}$) (dark green). The initial dimensionless coupling constants are $g_1=g_2=1.0,g_3=1.0, g_{u\parallel}=1.3, g_{u\perp}=1.0$ and $\alpha=0.8$. (a) and (b) show that the antiphase $d$-wave pairing is the leading instability.}
	\end{center}
	\label{scorder}
\end{figure}

In the RG approach, the pairing interactions are mediated by the electronic fluctuations.
{Fig.~5(a) shows that under the RG flow, the mediated pairing interaction in the antiphase $d$-wave pairing channel diverges faster than those in the CDW and SDW channels.} Different from a single idealized square FS favoring a leading SDW instability \cite{dzyaloshinskii}, the presence of the second square FS enables the inter-FS scattering described by $g_{u\parallel}$ and $g_{u\perp}$ {to greatly enhance $\Gamma_{dSC}^{\pm}$}, allowing the antiphase $d$-wave SC instability to win over the density wave instabilities as revealed in Fig.~5(a). In, Fig.~5(b) we compare the RG flow in the different pairing channels. Since $g_{u\parallel}>g_{u\perp}$ due to the large scattering density of states connecting the parallel sections in Fig.~4(a), the antiphase $d$-wave pairing ($\Gamma_{dSC}^\pm$) also wins over the in-phase $d$-wave and the $s$-wave pairings to become the leading SC instability as can be seen from the RG flows in Fig.~5(b). Thus the RG analysis shows that the ground state in the two-orbital model at $x=0.4$ is a two-band $d_\pm$-wave superconductor, in agreement with the strong-coupling results discussed above.

\subsection{Antiphase $s$-wave superconducting state near  Cu $d^8$ configuration}

Finally, we summarize our findings on the SC state of the two-band Hubbard model at even higher hole doping concentrations. To this end, the $\delta$-oxygen density is increased to $\delta=0.45$, corresponding to a hole doping $x=0.90$ currently unreachable in the high-pressure experiments on Ba$_2$CuO$_{3+\delta}$ , where the copper valence is close to the $d^8$ configuration. Carrying out the strong coupling renormalized meanfield theory, we found that the ground state changes to a different SC state of $A_{1g}$ symmetry with a two-band antiphase $s_\pm$ gap function. Indeed, this two-band nodeless SC state
has been proposed recently \cite{jiang} for the monolayer CuO$_2$ grown on Bi$_2$Sr$_2$CaCu$_2$O$_{8+\delta}$ substrate \cite{xue} as electron doped two-orbital AF Mott insulator associated with the Cu $d^8$ configuration.

We have subsequently carried out the weak-coupling RG analysis of the two-band Hubbard model close to the Cu $d^8$ configuration. At $x=0.9$, the electron-like FS around the $\Gamma$ point is comparable in size to the hole-like FS around the M point. As a result, the inter-FS particle-hole scattering increases significantly and plays a key role in realizing an emergent $s$-wave SC state with an $s_{\pm}$ pairing gap function, consistent with the prediction of the strong-coupling renormalized meanfield theory. This finding is in qualitative agreement with the SC state obtained using the weak-coupling random phase approximation (RPA) for a related two-band Hubbard mode at $x=0.9$  \cite{maier}. Moreover, the result of the RG analysis is analogous to that obtained in weak-coupling studies of the multi-orbital Fe-based superconductors \cite{mazin,frg,hirshfeld-scalapino,chubukov08}.

This prediction of nodeless antiphase $s$-wave superconductor near the Cu $d^8$ configuration in unconventional cuprates awaits future materials realization. High-pressure growth of cuprates in strong oxidation environment, such as the realization of Ba$_2$CuO$_{3+\delta}$, may present a promising path.

\section{summary and discussion}

We presented a theoretical description of the atomic and electronic structure, and the emergence of high-$T_c$ superconductivity in the newly discovered 73K unconventional cuprates Ba$_2$CuO$_{3+\delta}$ at $\delta=0.2$ \cite{jin}. The key difference to the conventional cuprates is the high pressure and oxygenation growth stabilized polycrystals that are isostructural to La214, but with an the inverted crystal field due to the compressed octahedron that liberates the \dz orbital in addition to the \dx orbitals, realizing a highly hole overdoped ($x=0.4$) $d^9$ configuration. This is complimentary to the monolayer CuO$_2$ grown on Bi$_2$Sr$_2$CaCu$_2$O$_{8+\delta}$ substrate \cite{xue}, where the liberation of the \dz orbital and nodeless antiphase $s$-wave superconductivity was argued to arise from the crystal field of the unbalanced octahedron and the heavy carrier doping through the interface carrier transfer that realize an unconventional state near the Cu $d^8$ configuration \cite{jiang}.

Constructing a minimal two-orbital Hubbard model using the DFT band structure, we studied the compressed Ba214 with inverted \dz and \dx orbitals both in the undoped case with 3 electrons and in the heavily overdoped region.
We found that the correlated electronic states can be described by a doped orbital-polarized Mott insulator and a novel two-band high-T$_c$ supercondcutor with $d_\pm$-wave gap functions emerges through the spin and orbital exchange interactions. The multiband and the orbital-dependent correlations are crucial for achieving the high-$T_c$ superconductivity observed at a high doping near $x=0.4$, in contrast to the conventional single-band cuprates.
The basic prediction of the \dz orbital libration and the two-band superconductivity should be amenable to experimental tests for two pairing gaps in the heat capacity, NMR, and tunneling measurements.

These central results on the two-band antiphase $d$-wave SC state
do not depend on specific realizations of the SC Ba214 regions in Ba$_2$CuO$_{3+\delta}$. Even if the entire sample were viewed as compressed Ba214 with randomly distributed oxygen vacancies in the plane, one would arrive at the same conclusion if
the disorder effects of the significant oxygen vacancies can be ignored. However, it is known that the high density of such in-plane oxygen vacancies can be destructive for superconductivity in the cuprates. The proposed structures in Figs.~1(a) and 1(c) have the advantage of avoiding significant oxygen vacancies in the superconducting Ba214 regions. 
Accordingly, Ba$_2$CuO$_{3+\delta}$ may exhibit features of
granular superconductivity \cite{deutscher,mason2012,mason2020}, with grain sizes larger than the short coherence length $\xi$. Taking the SC gap $\Delta \sim 25$meV, we estimate $\xi=\hbar v_F/\pi\Delta\sim1$nm using the renormalized bands. In this case, the Josephson-coupled SC grains stabilize a macroscopic phase-coherent SC state \cite{larkin,spivak}.
The significant flux penetration revealed by the magnetization measurements,
the $\sim30$\% SC volume fraction from the dc susceptibility and $\mu$SR,
and the broad specific heat anomaly in the current experiments \cite{jin} are consistent with this picture.
While further experimental and theoretical studies are necessary,
the present theory offers a conjecture that the 73K Ba$_2$CuO$_{3+\delta}$ highlights a class of unconventional cuprates, including the high-pressure grown 95K Sr$_2$CuO$_{3+\delta}$ and 84K  Cu$_{0.75}$Mo$_{0.25}$Sr$_2$YCu$_2$O$_{7.54}$ \cite{geballe,uchida,taked,geballe2016},
as a possible route toward higher $T_c$ by utilizing the electrons partially occupying both of the copper $e_g$ orbitals.

\section{acknowledgement}
We thank C.Q. Jin and S. Uchida for useful discussions. The work is supported in part by the Ministry of Science and Technology of China 973 program (No. 2017YFA0303100, No. 2015CB921300), National Science Foundation of China (Grant No. NSFC-1190020, 11534014, 11334012), and the Strategic Priority Research Program of CAS (Grant No. XDB07000000); and the U.S. Department of Energy, Basic Energy Sciences Grant No. DE-FG02-99ER45747 (K.J and Z.W). Z.W. thanks IOP and KITS of CAS, and Aspen Center for Physics for hospitality, and the support of NSF Grant No. PHY-1066293.

\appendix

\section{Density functional theory (DFT) calculations}

Our calculations are performed using density functional theory (DFT) employing the projector augmented wave (PAW) method encoded in the Vienna ab initio simulation package (VASP) \cite{Kresse1993,Kresse1996,Kresse1996b}. Generalized-gradient approximation (GGA) \cite{Perdew1996} for the exchange correlation functional is used. Throughout the work, the cutoff energy is set to be 500 eV for expanding the wave functions into plane-wave basis. In the calculation, the Brillouin zone (BZ) is sampled in the $k$ space within Monkhorst-Pack scheme\cite{Monkhorst1976}. On the basis of the equilibrium structure, the $k$ mesh used is $10\times10\times4$.
\begin{table}
	\caption{Experimental determined crystal Structure of Ba$_2$CuO$_{3+\delta}$ in space group $I4/mmm$ with lattice constant $a=4.003\AA$ and $b=12.942\AA$ \cite{jin}.\label{structure}}
	\begin{tabular}{c|c|c|c|c|c}
		\hline
		\hline
		Atom & site & x  & y & z & occupancy\\	\hline
		Ba & 4e & 0 & 0 & 0.35627 & 1 \\	\hline
		Cu & 2a & 0 & 0 & 0 & 1 \\	\hline
		O$_1$ & 4e & 0 & 0 & 0.1438 & 1\\	\hline
		O$_2$ & 4c & 0 & 0.5 & 0 & 0.592\\	\hline \hline
	\end{tabular}
\end{table}

In our DFT calculations, we adopt the experimental parameters listed in Table \ref{structure}, which are extracted from the supplemental section in Ref.~\cite{jin}, and use the stoichiometric formula Ba$_2$CuO$_{4}$ (Ba214) and  Ba$_2$CuO$_{3}$ (Ba213), whose crystal structures are shown in Fig.~6(a) and Fig.~7(a). The calculated band structures are shown in Fig.~6(c) and Fig.~7(c) for Ba214 and Ba213 respectively. Note that the convention of the BZ using the BCO primitive unit cell is slightly different than the normal cuprates convention. To be consistent with studies of other cuprates, we use the common conventions for the cuprates.
As can be seen from Fig.~7, Ba213 has one-dimensional (1D) Cu-O chain planes. We choose the Cu-O chain to be along the $x$ direction.
As a result, the hopping in the $y$ direction is greatly reduced due to the lack of the oxygen in the Cu-Cu bond, leading to an 1D band with two 1D Fermi surface sections shown in Fig.~7(b) and (c). These are consistent with the calculations in Ref.~\cite{taoxiang}.

As discussed in the main text, we also studied a bilayer structure with alternating Ba213 and Ba214 planes shown in Fig.~8(a). The orbital-resolved electronic band dispersions in the Ba214 Cu$O_2$ plane  are shown in Fig.~8(b), which are very similar to those of the compressed single-layer Ba214 (Fig.~6(c)). Moreover, the orbital-resolved band dispersions projected to the Ba213 Cu-O chain plane shown in Fig.~8(c) closely follow those of the bulk Ba213 (Fig.~7(c)).
\begin{figure}
	\begin{center}
		\fig{3.4in}{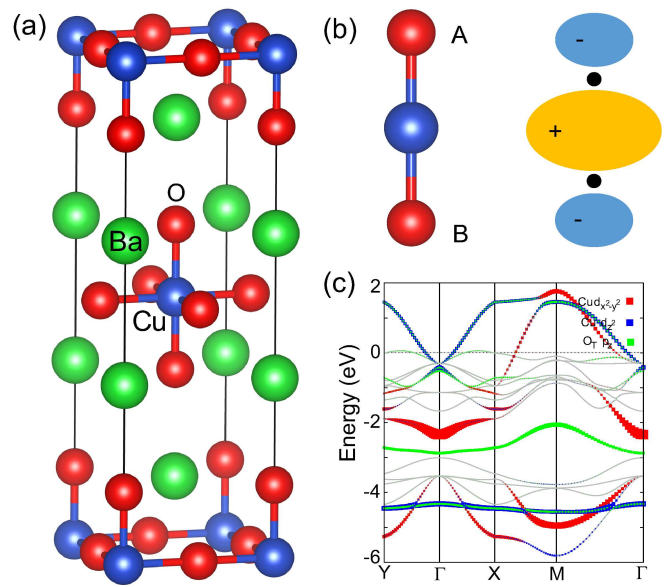}\caption{(a) Crystal structure of  Ba$_2$CuO$_{4}$.
			(b) Effective molecular orbital through hybridization between the Cu \dz orbital and apical oxygen p$_z$ orbital. Right side is the schematic molecular orbital $\phi_{1}$. (c) Orbital-resolved DFT band structure for compressed Ba214 calculated using the structural parameters in Table S1.}
	\end{center}
	\label{scorder}
\end{figure}

\section{Orbital construction and tight-binding model parameters}
To construct an effective model describing the band structures near Fermi level, we can analyze orbital characters of the electronic structure. The $d_{x^2-y^2}$ orbital mixed with the anti-symmetric combination of the in-plane oxygen p$_x$ and p$_y$ orbitals is similar to the Zhang-Rice singlet of common cuprate contributing a hole pocket around the M point. Due to the compressed octahedron, the $d_{z^2}$ orbital strongly hybridizes with the p$_z$ orbital from the top apical oxygen (O$_A$) and bottom apical oxygen (O$_B$), as shown in the left side of Fig.~6(b).

\begin{figure}
	\begin{center}
		\fig{3.4in}{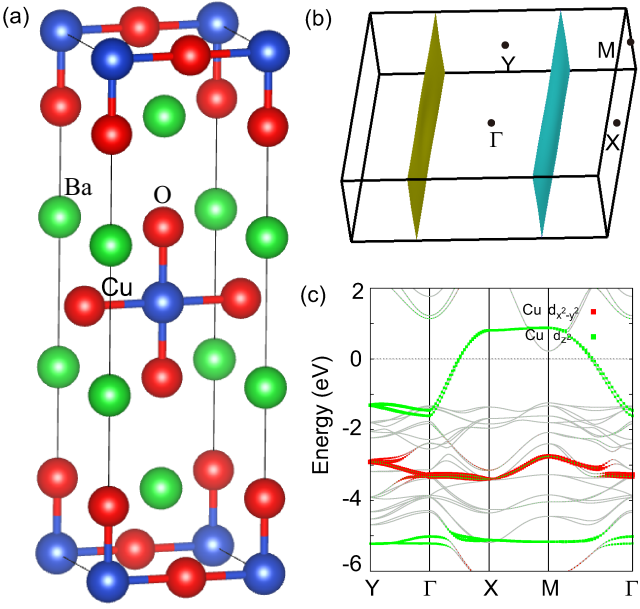}\caption{(a) Crystal structure of Ba$_2$CuO$_{3}$ with Cu-O chains along the $x$ direction. (b) The Fermi surface of Ba$_2$CuO$_{3}$.
			(c) Orbital-resolved DFT band dispersions.}
	\end{center}
	\vskip-0.5cm
	\label{scorder}
\end{figure}

\begin{figure}
	\begin{center}
		\fig{3.4in}{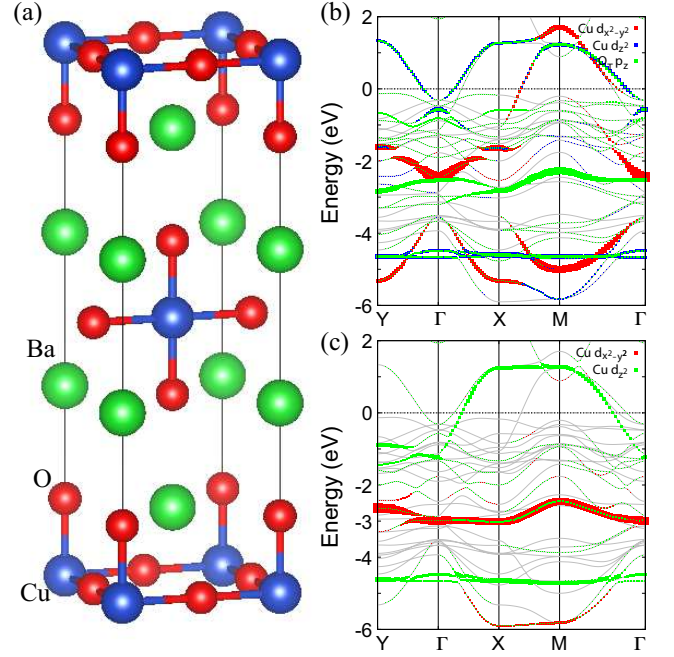}
		\caption{(a) Crystal structure of Ba213/Ba214 bilayers corresponding to Ba$_2$CuO$_{3.5}$ (b) Orbital-resolved band dispersions in the CuO$_2$ plane (Ba214).
			(c) Orbital-resolved band dispersions in the chain plane with Cu-O chains along $x$-direction (Ba213).}
	\end{center}
	\vskip-0.5cm
	\label{scorder}
\end{figure}
One can consider a local molecular model describing this hybridization.  Taking p$_z^A$,d$_{z^2}$,p$_z^B$ as the basis, the effective Hamiltonian of the molecular model can be written as
\begin{eqnarray}
	\label{local}
	H_{local}=\left(\begin{array}{ccc}
		\epsilon_{1}& t & 0 \\
		t    & \epsilon_{2} & -t   \\
		0    &   -t    &\epsilon_{1}   \\
	\end{array}\right) \\ \nonumber
\end{eqnarray}
where $\epsilon_{1}$ and $\epsilon_{2}$ are the on-site energies of the p$_z$ and d$_{z^2}$ orbitals. $t$ is the hopping parameter between p$_z$ and d$_{z^2}$ orbitals. The eigenvalues of Eq.~(\ref{local}) can be found as
\begin{eqnarray}
	\label{local_e}
	E_1 & = &\frac{1}{2}(\epsilon_{1}+\epsilon_{2}+\sqrt{8t^2+(\epsilon_{1}-\epsilon_{2})^2})\nonumber
	\\
	E_{2} & = &\epsilon_{1}\nonumber
	\\
	E_{3} & = &\frac{1}{2}(\epsilon_{1}+\epsilon_{2}-\sqrt{8t^2+(\epsilon_{1}-\epsilon_{2})^2}).
\end{eqnarray}
The corresponding eigenvectors are
\begin{eqnarray}
	\phi_1 & = &-p^A_z+p^B_z-\frac{1}{2t}(-\epsilon_{1}+\epsilon_{2}+\sqrt{8t^2+(\epsilon_{1}-\epsilon_{2})^2})d_{z^2}\nonumber
	\\
	\phi_{2} & = &p^A_z+p^B_z\nonumber
	\\
	\phi_{3} & = &-p^A_z+p^B_z-\frac{1}{2t}(-\epsilon_{1}+\epsilon_{2}-\sqrt{8t^2+(\epsilon_{1}-\epsilon_{2})^2})d_{z^2}.\nonumber
\end{eqnarray}
The schematic molecular orbital $\phi_{1}$ is plotted in right side of Fig.~6(b).
From Fig.~6(c), we can also find that $\phi_{1}$ and $\phi_{3}$ are located around Fermi level and -4.5eV respectively. $\phi_{2}$ is entirely attributed to O$_{A/B}$-p$_z$ orbitals and distributes around -2.6eV. Hence, $\phi_{1}$ with the \dz like bonding orbital and Zhang-Rice singlet with the \dx like bonding orbital dominate the electronic structure around the BCO Fermi level.

Based on the DFT results and orbital fields, we construct a two-orbital tight-binding (TB) model of Cu $e_g$ complex for the BCO. The Hamiltonian is given in Eq.~(1) in the main text. Denoting $d_{\alpha\sigma}$, $\alpha=x$($d_{x^2}$)$,z$($d_{z^2}$)
\begin{eqnarray}
	H_{t}&=&\sum_{k\sigma}\varepsilon_k^{xx}
	d_{kx\sigma}^\dagger d_{kx\sigma}
	+ \sum_{k\sigma}\varepsilon_k^{xz}
	(d_{kx\sigma}^\dagger d_{kz\sigma}+h.c.)
	\nonumber \\
	&+&\sum_{k\sigma}\varepsilon_k^{zz}
	d_{kz\sigma}^\dagger d_{kz\sigma}+ \sum_{k\sigma} e_\alpha d_{\alpha\sigma}^\dagger d_{\alpha\sigma}
	\label{Sht}
\end{eqnarray}
where $\varepsilon_k^{\alpha\beta}$ is the kinetic energy due to intra and interorbital hopping, and $e_\alpha$ is the on-site energy of \dz and \dx orbitals. Up to third nearest neighbor hopping, we have
\begin{eqnarray}
	\varepsilon_k^{xx}&=&-2t_x\gamma_k-4t_x^\prime \alpha_k-2t_x^{\prime\prime}\gamma^\prime_k
	\nonumber \\
	\varepsilon_k^{zz}&=&-2t_z \gamma_k-4t_z^\prime \alpha_k-2t_z^{\prime\prime}\gamma^\prime_k
	\nonumber \\
	\varepsilon_k^{xz}&=&2t_{xz} \beta_k+2t_{xz}^{\prime\prime}\beta^\prime_k
	\label{Sek}
\end{eqnarray}
where the intraorbital hopping involves lattice harmonics of $A_1$ symmetry $\gamma_k=\cos k_x+\cos k_y$, $\alpha_k=\cos k_x\cos k_y$, and $\gamma_k^\prime=\cos2k_x+\cos2k_y$, and the interorbital hopping involves $B_1$ harmonics $\beta_k=\cos k_x-\cos k_y$ and $\beta_k^{\prime}=\cos2 k_x-\cos 2k_y$.
The hopping parameters for the 1st ($t$), 2nd ($t^\prime$), and 3rd ($t^{\prime\prime}$) nearest neighbors are listed in Table II. The on-site energy of \dx and \dz are $e_{x}=2.275eV$ and $e_{z}=3.2035eV$.
\begin{table}[t]
	\caption{Hopping parameters of the TB model in eV.}
	
	\begin{tabular}{c|c|c|c}
		\hline
		\hline
		hopping integral & 1rd ($t$) & 2nd ($t^\prime$) & 3rd ($t^{\prime\prime}$)\\	\hline
		intra-orbital $t_x$ & -0.4968 & 0.0503 & -0.0652 \\	\hline
		intra-orbital $t_z$ &  -0.2135 & -0.0190 & -0.0219\\	\hline
		inter-orbital $t_{xz}$ & 0.3324 & 0.0 & 0.0390 \\	\hline \hline
	\end{tabular}
\end{table}

\end{document}